\documentclass[twocolumn]{aastex631}

\shorttitle{High-energy $\nu$ emission associated with gravitational waves }
\shortauthors{Matsui et al.}
\graphicspath{{./}{figures/}}
\usepackage{nccmath}
\newcommand{\tb}{Table }
\newcommand{\fg}{Figure }
\newcommand{\equ}{Eq.}

\begin{document}

\title{High-energy neutrino emission associated with gravitational-wave signals:\\
effects of cocoon photons and constraints on late-time emission}

\correspondingauthor{Riki Matsui}
\email{riki.matsui@astr.tohoku.ac.jp}

\author[0000-0003-0805-7741]{Riki Matsui}
\affiliation{Astronomical Institute, Graduate School of Science, Tohoku University, Sendai 980-8578, Japan}

\author[0000-0003-2579-7266]{Shigeo S. Kimura}
\affiliation{Frontier Research Institute for Interdisciplinary Sciences, Tohoku University, Sendai 980-8578, Japan}
\affiliation{Astronomical Institute, Graduate School of Science, Tohoku University, Sendai 980-8578, Japan}

\author[0000-0002-7114-6010]{Kenji Toma}
\affiliation{Frontier Research Institute for Interdisciplinary Sciences, Tohoku University, Sendai 980-8578, Japan}
\affiliation{Astronomical Institute, Graduate School of Science, Tohoku University, Sendai 980-8578, Japan}

\author[0000-0002-5358-5642]{Kohta Murase}
\affiliation{Department of Physics; Department of Astronomy and Astrophysics; Center for Multimessenger Astrophysics, Institute for Gravitation and the Cosmos, The Pennsylvania State University, University Park, Pennsylvania 16802, USA}
\affiliation{School of Natural Sciences, Institute for Advanced Study, }
\affiliation{Center for Gravitational Physics and Quantum Information, Yukawa Institute for Theoretical Physics, Kyoto University, Kyoto, Kyoto 606-8502, Japan}

%% Note that the \and command from previous versions of AASTeX is now
%% depreciated in this version as it is no longer necessary. AASTeX 
%% automatically takes care of all commas and "and"s between authors names.

%% AASTeX 6.31 has the new \collaboration and \nocollaboration commands to
%% provide the collaboration status of a group of authors. These commands 
%% can be used either before or after the list of corresponding authors. The
%% argument for \collaboration is the collaboration identifier. Authors are
%% encouraged to surround collaboration identifiers with ()s. The 
%% \nocollaboration command takes no argument and exists to indicate that
%% the nearby authors are not part of surrounding collaborations.

%% Mark off the abstract in the ``abstract'' environment. 
\begin{abstract}

We investigate prospects for the detection of high-energy neutrinos produced in the prolonged jets of short gamma-ray bursts (sGRBs). The X-ray lightcurves of sGRBs show extended emission components lasting for 100-1000 seconds, which are considered to be produced by prolonged engine activity. Jets by prolonged engine activity should interact with photons in the cocoon formed by the jet propagation inside the ejecta of neutron star mergers. We calculate neutrino emission from jets by prolonged engine activity, taking account of the interaction between photons provided from the cocoon and cosmic rays accelerated in the jets. We find that IceCube-Gen2, a future neutrino telescope, with the second-generation gravitational wave detectors will probably be able to observe neutrino signals associated with gravitational waves with around 10 years of operation, regardless of the assumed value of the Lorentz factor of the jets. 
Neutrino observations may enable us to constrain the dissipation region of the jets. We apply this model to GRB 211211A, a peculiar long GRB whose 
origin may be a binary neutron-star merger.
Our model predicts that IceCube is unlikely to detect any associated neutrino, but a few similar events will be able to put a meaningful constraint on the physical quantities of the prolonged engine activities.

\end{abstract}

%% Keywords should appear after the \end{abstract} command. 
%% The AAS Journals now uses Unified Astronomy Thesaurus concepts:
%% https://astrothesaurus.org
%% You will be asked to selected these concepts during the submission process
%% but this old "keyword" functionality is maintained in case authors want
%% to include these concepts in their preprints.

%% From the front matter, we move on to the body of the paper.
%% Sections are demarcated by \section and \subsection, respectively.
%% Observe the use of the LaTeX \label
%% command after the \subsection to give a symbolic KEY to the
%% subsection for cross-referencing in a \ref command.
%% You can use LaTeX's \ref and \label commands to keep track of
%% cross-references to sections, equations, tables, and figures.
%% That way, if you change the order of any elements, LaTeX will
%% automatically renumber them.
%%
%% We recommend that authors also use the natbib \citep
%% and \citet commands to identify citations.  The citations are
%% tied to the reference list via symbolic KEYs. The KEY corresponds
%% to the KEY in the \bibitem in the reference list below. 

\section{Introduction} \label{sec:intro}
Binary neutron star (BNS) mergers are one of the most important targets of gravitational-wave (GW) observations and electromagnetic (EM) follow-up observations.
Immediately after the first BNS merger event, GW170817, a short gamma-ray burst (sGRB) was identified as an EM counterpart \citep{AbbottA2017ApJ...848L..12A, AbbottB2017PhRvL.119p1101A, AbbottC2017ApJ...848L..13A}.
Radio observations show the superluminal motion \citep{MooleyA2018Natur.554..207M}, and
radio to X-ray counterparts are well modeled by off-axis afterglow emission from a structured relativistic jet \citep{MooleyA2018Natur.554..207M,MooleyB2018ApJ...868L..11M,TrojaA2018MNRAS.478L..18T,Ghirlanda2019Sci...363..968G,LambA2019ApJ...870L..15L}.
These studies support BNS mergers as the progenitor of sGRBs.

The formation mechanism and dissipation process of sGRB jets, however, are still unknown, despite a lot of theoretical and observational studies \citep{Berger2014}.
The standard afterglow scenario cannot explain typical X-ray lightcurves of sGRBs, where we see some excesses on a timescale of $10^2$-$10^5$ seconds after the prompt emission \citep{Norris2006ApJ...643..266N,Sakamoto2011ApJS..195....2S,Kagawa2015ApJ...811....4K, Kagawa2019ApJ...877..147K,Kaneko2015MNRAS.452..824K}.
These emission components, called extended and plateau emissions, are considered evidence of prolonged central engine activity \citep{Ioka2005ApJ...631..429I,Perna2006ApJ...636L..29P,Metzger2008MNRAS.385.1455M,Rowlinson2013MNRAS.430.1061R,Gompertz2014MNRAS.438..240G,Kisaka2015ApJ...804L..16K, Kisaka2017ApJ...846..142K}.
Current observations provide little constraint on the physical quantities of late-time emission components, including composition, Lorentz factor, and dissipation radius.

High-energy neutrinos are considered a powerful probe to investigate the physical quantities of gamma-ray bursts (GRBs) \citep{Waxman1997PhRvL..78.2292W,Guetta2004APh....20..429G,Murase2006PhRvD..73f3002M,Hummer2012PhRvL.108w1101H,Li2012PhRvD..85b7301L,He2012ApJ...752...29H,Kimura2017ApJ...848L...4K}. 
When jets dissipate their kinetic energy, electrons are non-thermally accelerated by some process, such as the first-order Fermi acceleration, and produce gamma-rays observed as GRBs.
If protons of PeV to EeV energies are accelerated at the same time, photohadronic interactions can produce neutrinos with energies above PeV \citep{Kimura2022arXiv220206480K}.
Observing such neutrinos in addition to EM wave signals enables us to investigate the physical mechanism of GRBs. 

The neutrino observatory, IceCube, has been detecting high-energy neutrinos from astrophysical objects and trying to determine their source for more than ten years \citep{Aartsen2013PhRvL.111b1103A,Aartsen2020PhRvL.124e1103A, ICCollaboration2021arXiv210109836I}.
Despite the expectation of neutrino emission from GRBs, GRB analyses by IceCube Collaboration revealed no significant spatial and temporal associations between cosmic neutrino events and GRBs \citep{Abbasi2010ApJ...710..346A,Abbasi2011PhRvL.106n1101A,ICCollaboration2012Natur.484..351I,Aartsen2015ApJ...805L...5A,Aartsen2016ApJ...824..115A,Aartsen2017ApJ...843..112A,Abbasi2022arXiv220511410A}, which put an upper limit on neutrino emission from GRBs.
Future cosmic high-energy neutrino detectors, such as IceCube-Gen2 \citep{Aartsen2021JPhG...48f0501A}, KM3Net/ARCA \citep{Aiello2019APh...111..100A}, baikal-GVD \citep{Avrorin2014NIMPA.742...82A}, P-One \citep{Agostini2020NatAs...4..913A}, and TRIDENT \citep{Ye2022arXiv220704519Y}, will significantly increase the detection rate of cosmic neutrinos.
We should estimate neutrino emissions from various environments to efficiently interpret the data. 

When we consider prolonged engine activity, the matter ejected by the BNS merger can play an important role in the high-energy component of late-time emissions.
BNS merger ejects outflowing material (ejecta), as confirmed by optical/infrared counterparts to GW170817 \citep{Kasliwal2017Sci...358.1559K,Kasen2017Natur.551...80K,Murguia2017ApJ...848L..34M,Shibata2017PhRvD..96l3012S,Tanaka2017PASJ...69..102T}.
The sGRB jet can interact with the ejecta if the jet formation delay from the merger, which leads to the formation of a cocoon \citep{Bromberg2011ApJ...740..100B, Hamidani2020MNRAS.491.3192H, Hamidani2021MNRAS.500..627H}.
The cocoon can provide photons into the dissipation region of prolonged jets. 
Leptonic emission considering the external photons from the cocoon has been discussed \citep{KimuraMurase+19,Toma+2009}, but hadronic emissions with the cocoon photons are not considered in detail.

In this study,  we calculate the production of high-energy neutrinos through the interaction of photons with cosmic rays accelerated inside the jet, taking into account the cocoon photons entering the prolonged jet.
We also calculate the possibility of detecting neutrinos at the late time of sGRBs associated with GWs, based on the sensitivities of IceCube and IceCube-Gen2 \citep{ICCollaboration2021arXiv210109836I}.
The calculation shows that after $\sim10$ years of observation by IceCube-Gen2, it is highly probable to observe one or more neutrinos in this scenario, or we can constrain the parameters with a 2 or 3 $\sigma$ confidence level. 
This result is almost independent of the Lorentz factor of jets, which may enable us to put a stronger constraint on the dissipation radius than that without cocoon photons. 
We describe our model and show the neutrino spectra in Section \ref{sec:model}. In Section \ref{sec:detection}, we discuss the probability of the neutrino detection with current and future detectors. In Section \ref{sec:GRB211211A}, we apply our model to GRB 211211A, a peculiar long GRB whose origin may be a BNS merger. 
Summary and discussion are described in Section \ref{sec:Summary}.
We use the notation $Q_X = Q/10^X$ in cgs unit unless otherwise noted and write $Q\prime$ for the physical quantities in the comoving frame of the
jet.

\section{Neutrino Production Using Cocoon photons}
\label{sec:model}

\subsection{Photon Distributions}
Neutrinos from GRBs are mainly produced by photomeson production. 
We consider that cosmic-ray (CR) protons are accelerated at a dissipation region in the jets. 
Photon distributions in the dissipation region affect the resulting neutrino spectra. Here, we consider two photon components: non-thermal radiation inside the jets and thermal radiation from the cocoon.

We write the differential number density of the non-thermal component using Band function:
\begin{eqnarray}
\frac{d n^{\prime \mathrm{in}}}{{d \varepsilon^{\prime}_\gamma}} = & n_{\varepsilon^{\prime}_\gamma,\mathrm{nor}} \left\{
\begin{array}{ll}
  \varepsilon^{\prime \alpha}_\gamma \ \mathrm{exp}\left(-\frac{(2+\alpha) \varepsilon^\prime_\gamma}{\varepsilon^{\prime}_{\gamma,\mathrm{pk}}}\right) & (\varepsilon^{\prime}_\gamma \leq \chi \varepsilon^{\prime}_{\gamma,\mathrm{pk}})\\
  \varepsilon^{\prime \beta}_\gamma \ (\chi \varepsilon^{\prime}_{\gamma,\mathrm{pk}}/e)^{\alpha-\beta} & (\varepsilon^{\prime}_{\gamma}>\chi \varepsilon^{\prime}_{\gamma,\mathrm{pk}})
\end{array}
\right.,
\label{internal}
\end{eqnarray}
where $\alpha$ and $\beta$ are the photon indices of X-rays below and above the peak, respectively, $\chi=(\alpha-\beta)/(2+\alpha)$, $n_{\varepsilon^{\prime}_\gamma,\mathrm{nor}}$ is the normalization factor, and $\varepsilon_\gamma^\prime$  and $\varepsilon^{\prime}_{\gamma,\mathrm{pk}}$ are the photon energy and the spectral peak energy in the jet comoving frame, respectively.
The normalization is determined so that $L_{\mathrm{X,iso}} =  4 \pi\Gamma_j^2 r_\mathrm{diss}^2 c \int^{\varepsilon_{\mathrm{X,Max}/\Gamma_j}}_{\varepsilon_{\mathrm{X,min}/\Gamma_j}} d\varepsilon^{\prime}_{\gamma} \varepsilon^{\prime}_{\gamma}({dn^{\prime}_\gamma}/{d\varepsilon^{\prime}_{\gamma}} )$ is satisfied, where $L_{X,\mathrm{iso}}$ is the isotropic-equivalent luminosity in the X-ray band, $r_\mathrm{diss}$ is the dissipation radius, $\Gamma_j$ is the Lorentz factor of the jets, and $\varepsilon_{\mathrm{X,Max}}$, and $\varepsilon_{\mathrm{X,min}}$ are the maximum and minimum energies of the X-ray band, respectively. 
Observationally, the X-ray luminosity shows a correlation with the duration of the extended emission as \citep{Kisaka2017ApJ...846..142K}
\begin{equation}
L_{X,\mathrm{iso}} = 2\times10^{49} \left(\frac{t_\mathrm{dur}}{10^2\rm~s}\right)^{-2.5}\rm erg~s^{-1}.\label{eq:Lx}
\end{equation}
We give $t_{\rm dur}$ as a parameter, and use the $L_X-t_{\rm dur}$ relation to estimate the photon number density and the jet power.
The total photon luminosity is obtained by $L_{\gamma,\mathrm{iso}} =  4 \pi\Gamma_j^2 r_\mathrm{diss}^2 c \int^{\varepsilon^{\prime}_{\gamma,\mathrm{Max}}}_{\varepsilon^{\prime}_{\gamma,\mathrm{min}}} d\varepsilon^{\prime}_{\gamma} \varepsilon^{\prime}_{\gamma}({dn^{\prime}_\gamma}/{d\varepsilon^{\prime}_{\gamma}}) $, where $\varepsilon^{\prime}_{\gamma,\mathrm{min}}$ and $\varepsilon^{\prime}_{\gamma,\mathrm{max}}$ are the minimum and maximum photon energies, respectively.
We set $\varepsilon^{\prime}_{\gamma,\mathrm{min}}= 0.1$ eV and $\varepsilon^{\prime}_{\gamma,\mathrm{Max}}= 10^6$ eV because the synchrotron self-absorption and the pair creation are effective below and above the energies, respectively \citep{Murase2006PhRvL..97e1101M}.
The values of $\varepsilon^{\prime}_{\gamma,\mathrm{min}}$ and $\varepsilon^{\prime}_{\gamma,\mathrm{max}}$ do not strongly affect the results of this paper. 

We use one-zone approximation for the cocoon as discussed in \cite{KimuraMurase+19}. 
The cocoon temperature is obtained as $ T_\mathrm{coc} = 
[3\mathcal{E}_\mathrm{coc}/(4 \pi R_\mathrm{coc}^3 a_\mathrm{rad})]^{1/4}$, where $a_\mathrm{rad}$, $R_\mathrm{coc} = 3.0\times 10^{12} \,t_{\mathrm{dur},2.5}$ cm, and $\mathcal{E}_\mathrm{coc}$ are the radiation constant, the cocoon radius, and the thermal energy of the cocoon, respectively.
$\mathcal{E}_\mathrm{coc}$ is obtained by $\mathcal{E}_\mathrm{coc} = \mathcal{E}_\mathrm{adi} + \mathcal{E}_\mathrm{rad}$, where $\mathcal{E}_\mathrm{adi} =  8.8\times 10^{44}  t_{\mathrm{dur},2.5}^{-1} $ erg and $ \mathcal{E}_\mathrm{rad} = 9.3 \times 10^{44}\, t_{\mathrm{dur},2.5}^{-0.3}$ erg represent the contributions by the initial thermal energy of ejecta and by the radioactive decay of neutron-rich nuclei in the ejecta.

The photons in the cocoon need to diffuse into the jet to work as an external photon field. Taking this effect into account, the spectrum of the thermal external photons entering into the dissipation region from the cocoon is given by 
\begin{equation}
\varepsilon^{\prime}_\gamma \frac{d n^{\prime \mathrm{ex}}}{{d \varepsilon^{\prime}_\gamma}}  = \Gamma_j \frac{8 \pi (\varepsilon^{\prime}_\gamma/\Gamma_j)^3}{h^3c^3} \frac{1}{\mathrm{exp}  [\varepsilon^{\prime}_\gamma/(\Gamma_j k_B T_{\rm coc})]-1} \times e^{-\tau},
\end{equation}
where $ T_\mathrm{coc}$ is the cocoon temperature, $\tau$ is the lateral optical depth of the jet, respectively, $h$ is the Planck constant, $k_B$ is the Boltzmann constant, and $c$ is the speed of light. Hereafter, we call this component cocoon photons.
The lateral optical depth is estimated to be $\tau = L_{k,\mathrm{iso}} \sigma_T \theta_j/(4\pi r_\mathrm{diss}\Gamma_j^2m_p c^3) = 3.2 \  L_{k,\mathrm{iso},50.5} r_{\mathrm{dis},12}^{-1}\Gamma_{j,2}^{-2}(\theta_{j}/5^\circ)$, where $\sigma_T$, $\theta_j$, and $L_{k,\mathrm{iso}}$ are Thomson cross-section, opening angle of the jet and isotropic-equivalent kinetic luminosity, respectively.
$L_{k,\mathrm{iso}}$ is estimated to be $L_{k,\mathrm{iso}}= L_{p,\rm iso}/\epsilon_p = \xi_p L_{\gamma,\mathrm{iso}}/\epsilon_p = 30\  (\xi_p/10)(\epsilon_p/0.33)^{-1} L_{\gamma,\mathrm{iso}}$, where $L_{p,\rm iso}$ is the luminosity of accelerated protons, and $\xi_p$ and $\epsilon_p$ are parameters.

We ignore the cocoon photons for $R_\mathrm{coc} < r_\mathrm{diss}$ because the cocoon does not cover the dissipation region for such a condition.
This is a rough approximation because the number of cocoon photons provided in the dissipation region should change smoothly, but it does not significantly affect the results because $R_{\rm coc}>r_{\rm diss}$ is satisfied in most cases in this study.

Recently, \cite{Hamidani2022arXiv221000814H} showed that a large fraction of cocoon is confined in the ejecta. 
If we apply the simulation to our model, 
$\mathcal{E}_\mathrm{coc}$ can be 200 times lower, and the temperature and number density of the cocoon photons can decrease by a factor of $(1/100)^{1/4}$ and $(1/100)^{3/4}$ , respectively.
However, this effect does not change the results dramatically because the  protons 
lose all the energies owing to a high cocoon photon density even for such a low temperature (see \equ\ref{eq;a4}).

\subsection{Neutrino Spectra}
\label{Calculating the Spectrum}
Particle acceleration processes occurring in astrophysical environments usually lead to a power-law distribution function \citep[e.g.,][]{Blandford1987,Guo2020}, and we represent the cosmic-ray (CR) distribution function in the engine-rest frame as 
\begin{equation}
\frac{dN_p}{d\varepsilon_{p}} = N_{\varepsilon_p,\mathrm{nor}} \left(\frac{\varepsilon_p}{\varepsilon_{p,\mathrm{cut}}}\right)^{-p_\mathrm{inj}} \mathrm{exp}\left(-\frac{\varepsilon_p}{\varepsilon_{p,\mathrm{cut}}}\right),
\end{equation}
where $\varepsilon_{p,\rm cut}$ is the proton cutoff energy and $N_{\varepsilon_p,\mathrm{nor}}$ is the normalization factor.  
$N_{\varepsilon_p,\rm nor}$ is determined by $L_{p,\mathrm{iso}} t_\mathrm{dur}= \xi_p L_{\gamma,\mathrm{iso}} t_\mathrm{dur} =  \int^{\infty}_{\varepsilon_\mathrm{min}} d\varepsilon_{p} \varepsilon_{p}({dN_p}/{d\varepsilon_{p}}) $.
The cutoff energy is determined by the balance between acceleration and cooling timescales. 
We use $\varepsilon_{p,\rm min}=\Gamma_j\varepsilon^\prime_\mathrm{min}= 3 \Gamma_jm_pc^2$ as the minimum energy of cosmic-ray protons, and $\xi_p$ is the cosmic-ray loading parameter \citep{Murase2006PhRvD..73f3002M}.

The acceleration timescale is estimated to be $t^\prime_\mathrm{acc} =\varepsilon^\prime_{p}/(ceB^\prime)$, where  $ B^\prime = \sqrt{2L_{\gamma,\mathrm{iso}}\xi_B/(c \Gamma_j^2 r_\mathrm{diss}^2)}$ is magnetic field in the comoving frame.

The cooling rate is given by $t^{\prime-1}_\mathrm{cool} = t^{\prime-1}_\mathrm{ad}+t^{\prime-1}_\mathrm{BH}+t^{\prime-1}_{p\gamma} + t^{\prime-1}_{\mathrm{syn}} $, where each term represents adiabatic cooling, Bethe-Heitler process, photomeson production, and synchrotron cooling, respectively.  
The adiabatic and the synchrotron cooling timescales are given by $t^{\prime}_\mathrm{ad} = r_\mathrm{diss}/(\Gamma_j c)$ and $t^{\prime}_\mathrm{syn} = 6 \pi m_\mathrm{\rm{p}}^4c^3/(m_\mathrm{e}^2\sigma_T B^{\prime2})$, respectively.
The cooling rates for Bethe-Heitler and photomeson production processes are estimated to be
\begin{equation}
t^{\prime-1}_{p\gamma/\mathrm{BH}} = \frac{c}{2\gamma^{\prime2}_{p}}  \int^{\infty}_{\bar{\varepsilon}_\mathrm{th}} d\bar{\varepsilon}_\gamma \sigma(\bar{\varepsilon}_\gamma) \kappa(\bar{\varepsilon}_\gamma) \bar{\varepsilon}_\gamma \int^{\infty}_{\bar{\varepsilon}_\gamma/{2\gamma_p}} d\varepsilon^{\prime}_\gamma \varepsilon_\gamma^{{\prime}-2}\frac{dn^{\prime}_\gamma}{d\varepsilon^{\prime}_\gamma},
\label{pg}
\end{equation}
where $\gamma_p^\prime = \varepsilon^\prime_\mathrm{p}/(m_\mathrm{p} c^2)$ is the Lorentz factor of protons and $\bar{\varepsilon}_\mathrm{th}$, $\sigma(\bar{\varepsilon}_\gamma)$ and $\kappa(\bar{\varepsilon}_\gamma)$ are the threshold energy, the cross-section, and inelasticity for each reaction in the proton rest frame, respectively.
We use the fitting formulae based on GEANT4 for the cross-section and inelasticity for photomeson production \citep{Murase2006PhRvD..73f3002M} and analytic fitting formulae given in \cite{Stepney1983MNRAS.204.1269S, Chodorowski1992ApJ...400..181C} for Bethe-Heitler process.
We define $t_{p\gamma, \rm int}$, $t_{\rm BH,int}$, $t_{p\gamma,\rm int}$, and $t_{\rm BH,coc}$ as the cooling timescales using the internal photons and the cocoon photons, respectively.

The neutrino spectrum produced by the photomeson process and pion decay is approximately estimated to be
\begin{equation}
\frac{dN_{\nu_\mu}}{d\varepsilon_{\nu_\mu}} \approx \frac{25}{2} \int d\varepsilon_{\pi} g(\varepsilon_{\pi},\varepsilon_{\nu_\mu})f_{\mathrm{sup},\pi} \left.\left(f_{p\gamma} \frac{dN_p}{d\varepsilon_{p}}\right)\right|_{\varepsilon_{p} = 5\varepsilon_{\pi}}
\label{phinumu}
\end{equation}
for muon neutrinos, and 

\begin{equation}
\begin{split}
\frac{dN_{\bar{\nu}_\mu}}{d\varepsilon_{\bar{\nu}_\mu}} 
    &\approx \frac{dN_{\nu_e}}{d\varepsilon_{\nu_e}} \\
    &\approx \frac{50}{3}  \int d\varepsilon_{\mu} g(\varepsilon_{\mu},\varepsilon_{\nu_e}) f_{{\rm sup},\mu} \left. \left(f_{{\rm sup},\pi}f_{p\gamma} \frac{dN_p}{d\varepsilon_{p}}\right)\right|_{\varepsilon_{p} = 5\varepsilon_{\pi} = \frac{20}{3}\varepsilon_{\mu}}
\end{split}
\label{phinue}
\end{equation}
for anti-muon neutrinos and electron neutrinos, where 
$f_{p\gamma} = t^{\prime}_\mathrm{cool}/t^{\prime}_{p\gamma}$, $f_{{\rm sup},i}$, and $g(\varepsilon_{i},\varepsilon_{j})d\varepsilon_{j}$ are the pion production efficiency by photomeson production, the suppression factor by the pion and muon coolings, and the distribution function of the secondary particle $j$ produced by the decay of the parent particle $i$ of energy $\varepsilon_{i}$, respectively.
The suppression factor is estimated to be $f_{{\rm sup},i} = 1- \mathrm{exp}(-t^{\prime}_{i,\mathrm{cool}}/t^{\prime}_{i,\rm dec})$, where $t^{\prime}_{i,\rm dec}$ and $t^{\prime}_{i,\mathrm{cool}}$ are the lifetime and the cooling timescale of each particle in the comoving frame of the jet, respectively.
The lifetime is given by $t^{\prime}_{i,\rm dec} = t_i \varepsilon^\prime_i/(m_ic^2)$, where $t_i$ is the lifetime in the particle rest frame, and $m_i$ is the mass of a particle.
$t^{\prime}_{i,\mathrm{cool}}$ is estimated to be
$t^{\prime-1}_{i,\mathrm{cool}} = t^{\prime-1}_{i,\mathrm{syn}} +t^{\prime-1}_\mathrm{ad} $,
where
$t^{\prime}_{i,\mathrm{syn}} = 6 \pi m_i^4c^3/m_\mathrm{e}^2\sigma_T B^2 \varepsilon^\prime_i$. 
Here, we assume that all pions produced by the photomeson production with $\varepsilon_p$ have $\varepsilon_\pi =0.2\varepsilon_p$, and all muon produced by the decay of pions with $\varepsilon_\pi$ have $\varepsilon_\mu =(3/4) \varepsilon_\pi$.
We approximate $g(\varepsilon_{\pi},\varepsilon_{\nu_\mu}) = 4\Theta({\varepsilon_{\pi}}-4\varepsilon_{\nu_\mu})/\varepsilon_{\pi}$ and $g(\varepsilon_{\mu},\varepsilon_{\nu_e}) = 3\Theta(\varepsilon_{\mu}-3\varepsilon_{\nu_e})/\varepsilon_{\mu}$, where $\Theta(x)$ is Heaviside step function, because they imitate energy distributions for the two-body decay.
This treatment can approximately account for the low-energy tail of the neutrino spectrum, which would affect the detectability of neutrinos.

Taking account of the neutrino mixing, we can approximately obtain the neutrino fluences measured at the Earth as \citep[e.g.,][]{Becker2008}
\begin{equation}
\phi_{\nu_e+\bar{\nu}_e} \approx \frac{10}{18}  \phi_{\nu_e+\bar{\nu}_e}^0 +\frac{4}{18}  (\phi_{\nu_\mu+\bar{\nu}_\mu} ^0+ \phi_{\nu_\tau+\bar{\nu}_\tau}^0 )
\end{equation}
\begin{equation}
\phi_{\nu_\mu+\bar{\nu}_\mu}  \approx \frac{4}{18}  \phi_{\nu_e+\bar{\nu}_e}^0 +\frac{7}{18}  (\phi_{\nu_\mu+\bar{\nu}_\mu} ^0+ \phi_{\nu_\tau+\bar{\nu}_\tau}^0),
\end{equation}
where $\phi_{i}^0 = ({dN_{i}}/{d\varepsilon_{i}})/(4\pi d_L^2)$ is the neutrino fluence measured on the Earth assuming that the flavor ratio is fixed at the source, and $d_L$ is the luminosity distance.  %the source  

\begin{figure*}
\gridline{\fig{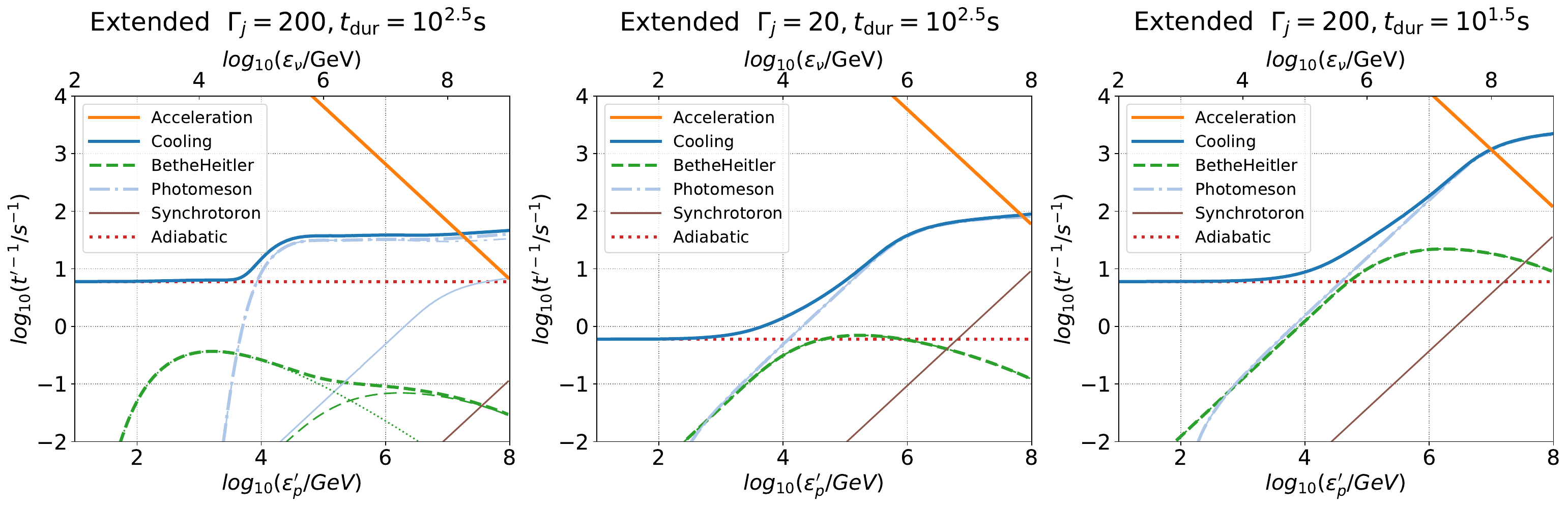}{1\textwidth}{}
          }
\gridline{\fig{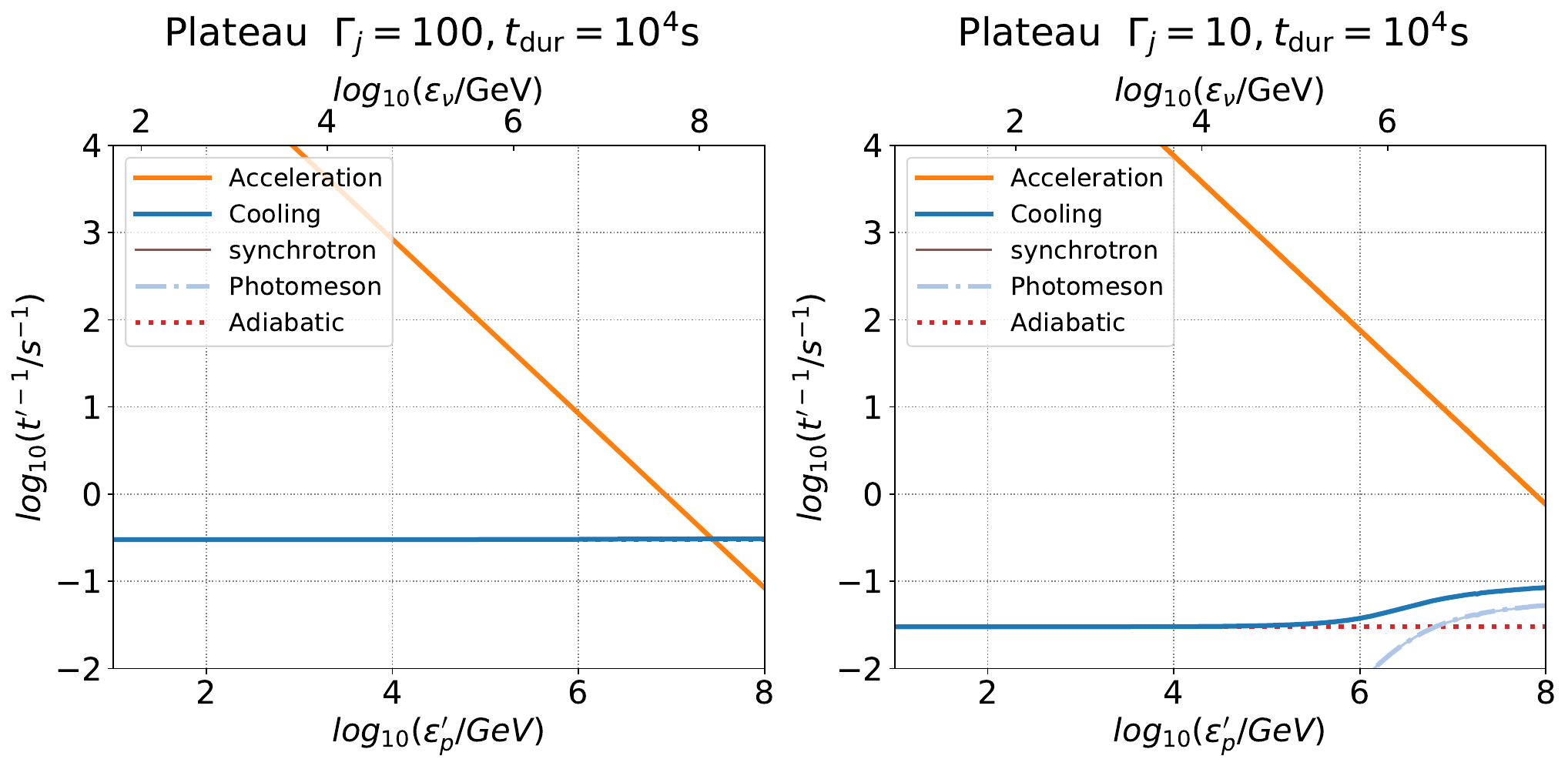}{0.66\textwidth}{}
          }
\caption{Acceleration and cooling rates for protons at the dissipation region in the comoving frame of the jet with various parameter sets. The thick orange lines and thick blue lines represent the acceleration timescales and total cooling timescales, respectively. The red thick-dotted, the lightblue-thick-dot-dashed, and the green-thick-dashed lines represent $t^{\prime-1}_{\rm ad}$, $t^{\prime-1}_{p\gamma}$, and $t^{\prime-1}_{\rm BH}$, respectively.
The lightblue-thin-solid and the lightblue-thin-dot-dashed lines represent $t^{\prime-1}_{p\gamma,\mathrm{int}}$ and $t^{\prime-1}_{p\gamma,\mathrm{coc}}$, respectively. The green-thin-dashed and the green-thin-dotted lines represent $t^{\prime-1}_{\mathrm{BH,int}}$ and $t^{\prime-1}_{\mathrm{BH,coc}}$, respectively.
}
\label{fig:tp-e}
\end{figure*}

\fg\ref{fig:tp-e} shows the acceleration and cooling timescales of protons as a function of energy (see \tb\ref{fiducial parameters} for our fiducial parameter set for extended and plateau emission). 
For our extended emission models, the photomeson production is the most efficient cooling process for $\varepsilon_p^{\prime}\gtrsim 10-100$ TeV whereas the adiabatic loss is the most efficient for $\varepsilon_p^{\prime}\lesssim10-100$ TeV.
The Bethe-Heitler process is not effective for any parameters due to its relatively low effective cross-section.
The synchrotron cooling is also not effective because of the heavy mass of a proton and the moderate magnetic field.
For the plateau emission models, the adiabatic loss is the most efficient except for $\varepsilon^\prime_p\gtrsim10$ PeV for $\Gamma_j=10$. This is because their lower cocoon photon density.

The fluences of $\nu_\mu+\bar{\nu}_\mu$ for the fiducial parameters and for some other parameters are shown in \fg\ref{fig:phi-e}.
In Appendix \ref{App_a}, we show analytic expression of the fluence for the extended emission model with fiducial parameters (blue lines), which is dominated by cocoon photons. We also show the parameter dependence there.

For the cases with the other parameter sets
shown in the panels (a) and (b) in \fg\ref{fig:phi-e}, the neutrinos are mainly produced by interaction with the internal photons.
For $\Gamma_j = 20$, $t_\mathrm{dur} = 10^{2.5}$, $\tau\sim100$ is achieved, and the cocoon photons cannot diffuse into the dissipation region in such a high optical depth. 
For $\Gamma_j = 200$, $t_\mathrm{dur} = 10^{1.5}$ s, 
$R_\mathrm{coc} < r_\mathrm{diss}$ is satisfied, and the cocoon photons cannot contribute to the neutrino emission. 
The neutrino spectra dominated by internal photons are roughly expressed by broken power law shapes with exponential cutoff that reflects spectra of internal photons and protons.

The parameter dependence of these neutrino spectra is consistent with previous studies that consider only internal photons \citep{Waxman1998PhRvD..59b3002W, Zhang2013PhRvL.110l1101Z,Kimura2017ApJ...848L...4K, Kimura2022arXiv220206480K}.
The neutrino spectra exhibit two break points: the low-energy break due to the photon spectral break and the high-energy break due to the pion cooling.
The relativistic beaming effect leads to $t_{\rm ad}^{-1}\propto \Gamma_j$ and $t_{p\gamma,\rm int}^{-1}\propto \Gamma_j^{-1}$, which can be seen in \fg\ref{fig:tp-e} by comparing the top-left and top-middle panels.
In this case, the neutrino fluence is written as $\phi_{\nu_\mu}\propto t_{p\gamma,\rm int}^{-1}/t_{\rm ad}^{-1}\propto \Gamma_j^{-2}$.
We can see this dependence by comparing the peaks by internal photons for $\Gamma_j = 20$ (orange dashed line) and for $\Gamma_j = 200$ (blue thin dashdot line) in panel (a) of \fg\ref{fig:phi-e}.
For the case with $\Gamma_j = $200 and $t_\mathrm{dur} = 10^{1.5}$ s, $t_{p\gamma}^{-1}$ and the fluence of neutrino is much higher than those for other parameters, 
because we set the luminosity $10^{2.5}$ times higher than that for fiducial parameters based on Eq. (\ref{eq:Lx}).

\begin{table*}
  \caption{fiducial parameters}
  %\centering
  \begin{tabular}{lcccccccc}
    \hline \hline
    Parameters  & $\Gamma_j$ & $t_\mathrm{dur}$ & $L_\mathrm{X,iso}$ &  $r_\mathrm{diss}$ & $\varepsilon_{\gamma,\mathrm{pk}}$  \\
     &  & (s) & (erg/s)  &  (cm) & (keV)  \\
    \hline 
    Extended  &  200  & $10^{2.5}$ &$10^{48}$ & $10^{12}$ & 10 \\
    Plateau  &  100  & $10^{4}$ & $10^{46}$ & $10^{13}$ & 1   \\
    \hline \hline 
    Shared  & $\alpha$ & $\beta$ & $p_\mathrm{inj}$ & $\xi_p$ & $\xi_B$ & $\varepsilon_{\mathrm{X,min}}$, $\varepsilon_{\mathrm{X,max}}$&$d_L $\\
     & & & & &  & (keV) &(Mpc)\\
    \hline 
     & $-0.5$ & $-2$ & 2.0 & 10 & 0.33 &  0.3 ,  10 (XRT) & 300\\
    \hline 
  \label{fiducial parameters}
  \end{tabular}
\end{table*}

\begin{figure}
\gridline{\fig{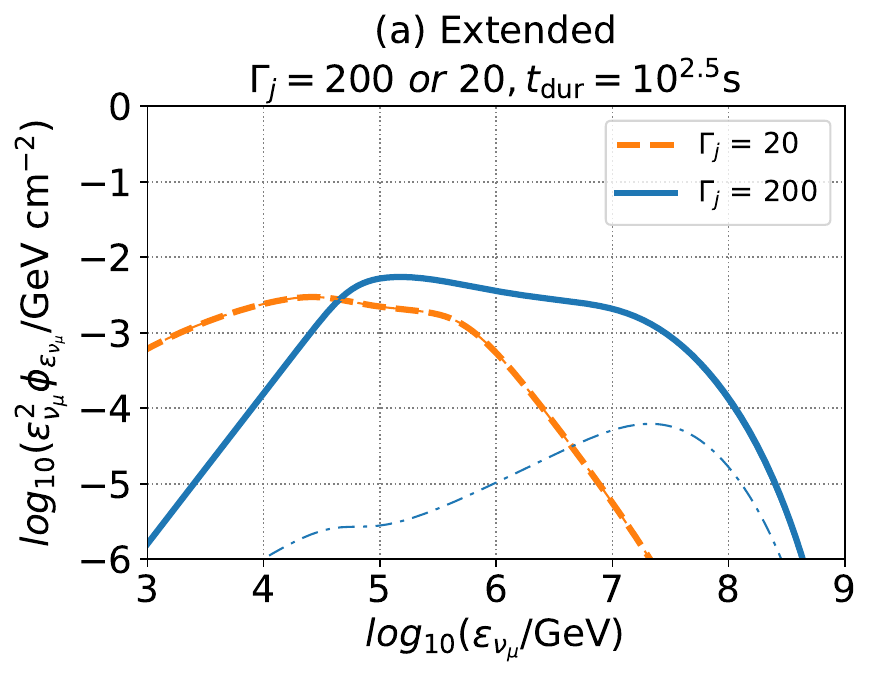}{0.38\textwidth}{}
          }
\gridline{\fig{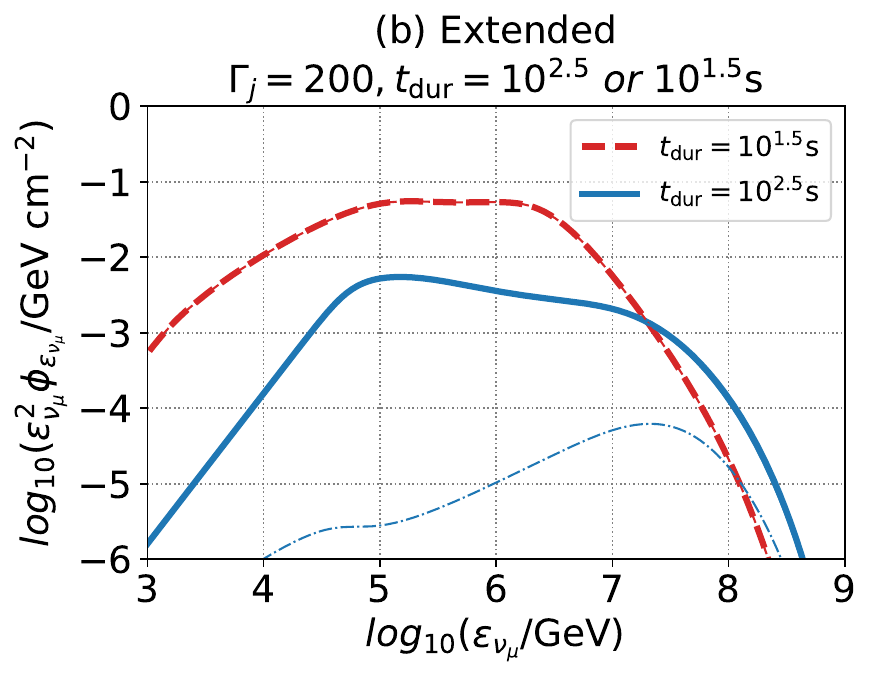}{0.38\textwidth}{}
}
\gridline{\fig{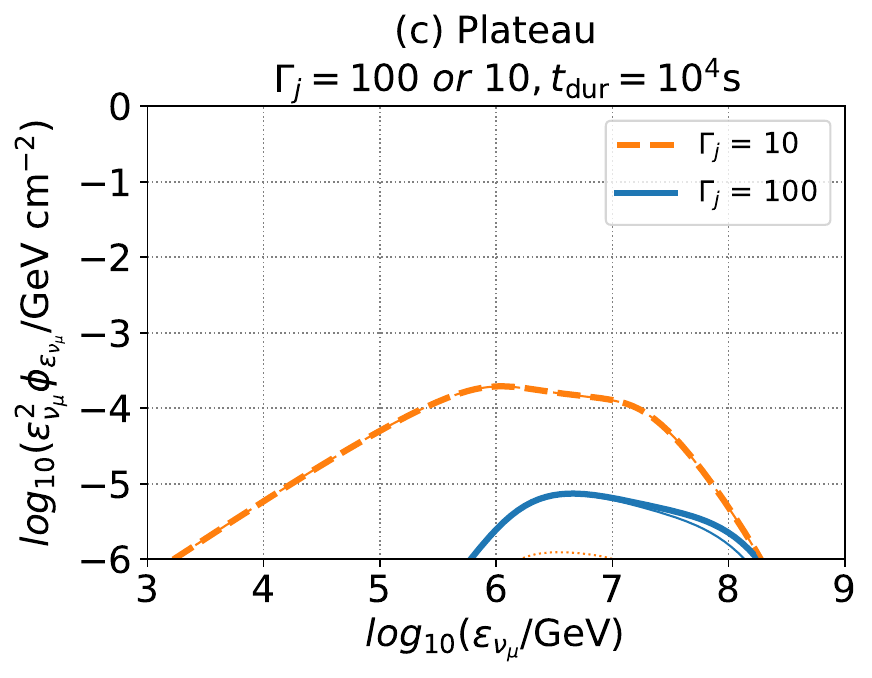}{0.38\textwidth}{}
}
\caption{Neutrino fluences at the Earth. The thick lines are the total fluence and the thin lines are the contributions by the internal photons (in panels (a) and (b)) or cocoon photons (panel (c)).%either the cocoon or internal photons. 
\ (a) Neutrino spectra for the extended emission models. The blue lines are for our fiducial parameter set, while the orange lines are for $\Gamma_j=20$.  The blue-thin-dot-dashed line is the contribution by the internal photons.
(b) Neutrino spectra for extended emission model. The blue lines are the same as (a). The red lines are for $t_\mathrm{dur} = 10^{1.5}$ s.
(c) Neutrino spectra for plateau emission models. The blue lines are for our fididucial parameters , while the orange line is for $\Gamma_j=10$. 
\label{fig:phi-e}}
\end{figure}

Neutrino fluences from plateau emissions are much lower than those for extended emissions. The difference between extended and plateau emissions is caused by the difference in the luminosity and the duration of the jet (see \tb\ref{fiducial parameters} for difference between extended and plateau emission).
For the plateau emission, the total energy of injected protons, $L_{\gamma,\rm{iso}} t_\mathrm{dur}$ are almost the same as that for extended emission.
On the other hand, photon number density for plateau emission is much less than that for extended emission because the longer duration leads to a low temperature cocoon due to the expansion. 
This results in a low cocoon photon number density at the dissipation region, causing a lower neutrino production rate for plateau emissions.

\section{Detection Prospects}
\label{sec:detection}
We discuss prospects for neutrino detection associated with gravitational waves.
Neutrino spectra from the extended emissions have a peak of $\sim 5.4 \times 10^{-3}$ GeV $\rm{cm}^{-2}$ at $\varepsilon_{\nu} \sim$ PeV, which is comparable to the design sensitivity for a transient object for IceCube-Gen2 \citep{Aartsen2021JPhG...48f0501A}.
These neutrinos should be detectable if we stack multiple sGRBs with extended emission, which requires long-term operation.
Since neutrinos from plateau emissions are too weak to be detected by near-future experiments, the following discussion focuses on the detectability of neutrinos from extended emissions.

\subsection{Neutrino Detection from a sGRB}
The expected number of $\nu_\mu$-induced events is estimated to be 
\begin{equation}
\bar{N}_{\nu_\mu} = \int d \varepsilon_{\nu_\mu} \phi_{\nu_\mu+\bar{\nu}_\mu} (\varepsilon_{\nu_\mu}) A_\mathrm{eff}(\delta,\varepsilon_{\nu_\mu}), 
\end{equation}
where $A_\mathrm{eff}$ is effective area for a detector 
and $\delta$ is declination angle. 
We use $A_{\rm eff}$ given in the 10-year point-source analysis \citep{ICCollaboration2021arXiv210109836I}, because this has a finer grid in $\delta$ than that used in the GRB analyses \citep{Aartsen2017ApJ...843..112A}.
We assume that the effective area of IceCube-Gen2 is 5 times larger than that of IceCube.

\begin{figure}
 \centering
 \includegraphics[width =0.49\textwidth]{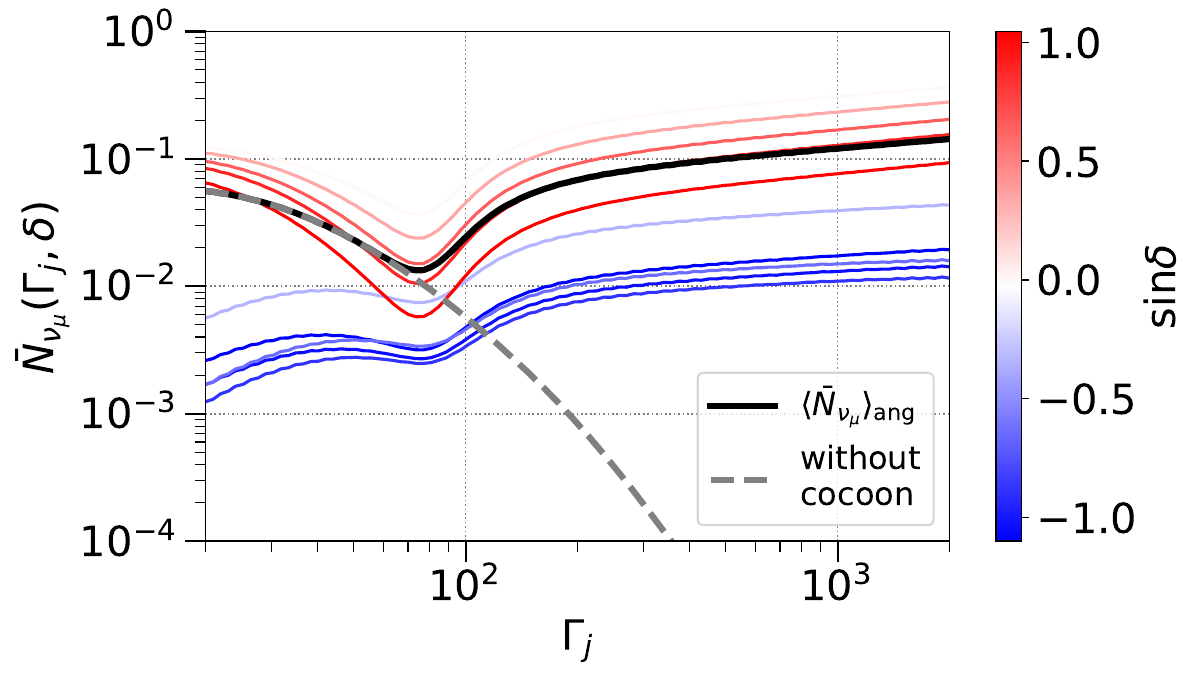}
 \caption{The expected number of $\nu_\mu$-induced events as a function of $\Gamma_j$ (colored-thin lines). The color bar represents the declination angle. The black-solid line represents the expected number averaged over the solid angle. The gray-dashed line represents the expected number averaged over the solid angle without the cocoon photons. \label{fig:N-Gamma}}
\end{figure}

We estimate $\bar{N}_{\nu_\mu}$ for various values of $\Gamma_j$ and $\delta$, whose results are shown in \fg\ref{fig:N-Gamma}.
We use the fiducial parameters for other parameters, including
$d_L =300$ Mpc, as shown in \tb\ref{fiducial parameters}.
The gray dashed line 
represents the expected number averaged over the solid angle without the cocoon photons. 
This line is proportional to $\Gamma_j^{-2}$ for $\Gamma_j< 100$ due to Lorentz beeming effect,
and to $\Gamma_j^{-4}$ for $\Gamma_j> 100$ additionally because the low-energy break is higher than the energy range where IceCube is sensitive.
The black solid line represents the expected number averaged over the solid angle with the cocoon photons. 
This line is overlapped with that without the cocoon photons for $\Gamma_j< 100$, since cocoon photons are negligible due to the shielding caused by large values of $\tau$.
For $\Gamma_j>100$, the two lines separate from each other, due to the contribution of the cocoon photons.
We find that if we take the cocoon photons into account, $\bar{N}_{\nu_\mu}$ has a relatively weak dependence on $\Gamma_j$, because the neutrino peak fluence does not strongly depend on $\Gamma_j$ when the cocoon photons are dominant, as shown in Appendix \ref{App_a}.
The expedted number is slightly higher for a higher $\Gamma_j$ as the muon and pion coolings are inefficient for a higher $\Gamma_j$.

\begin{figure}
\gridline{\fig{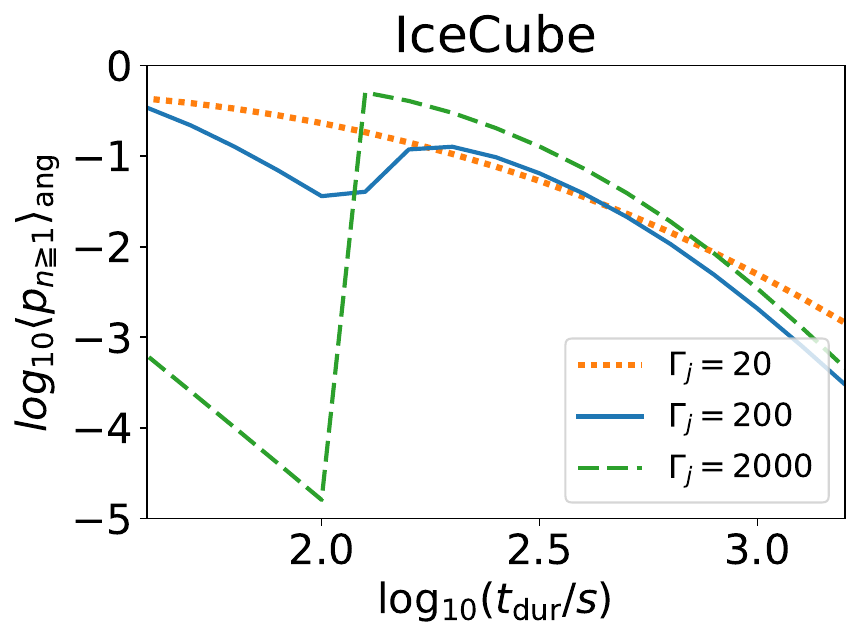}{0.4\textwidth}{}
}
\gridline{\fig{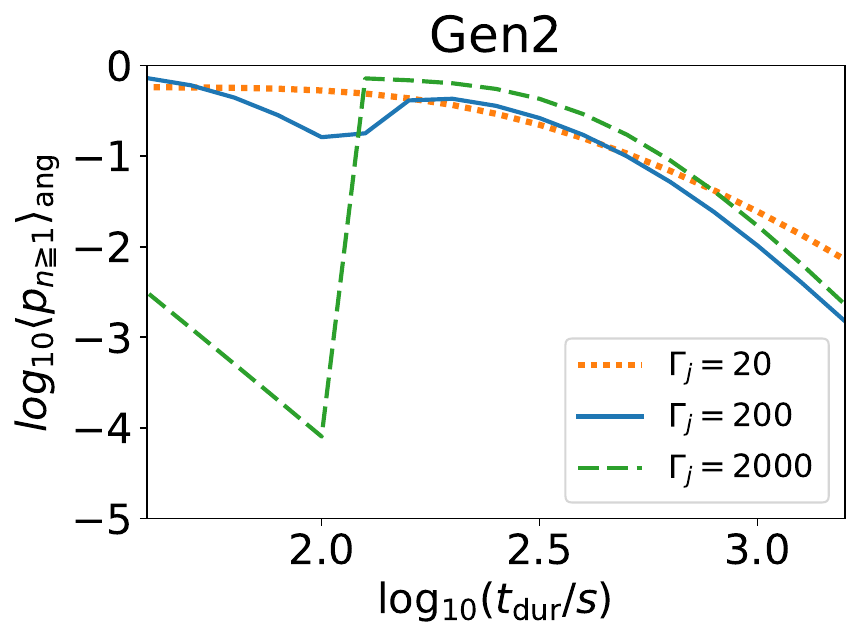}{0.4\textwidth}{}
          }
\caption{The probability of detecting more than 1 neutrinos as a function of $t_\mathrm{dur}$. The upper and lower panel are the models for IceCube and IceCube-Gen2, respectively. The dotted, solid, and dashed lines represent the probability for $\Gamma_j = 20$, 200, and 2000, respectively.     
\label{fig:p-t}}
\end{figure}

The dependence on the declination angle is caused by the effective area of IceCube.
IceCube efficiently detects neutrinos from the equatorial plane.
In the southern hemisphere, the low-energy neutrinos cannot be detected due to the atmospheric muons.
In the northern hemisphere, the high-energy neutrinos are absorbed by the Earth, which reduces the neutrino detection rate.  
For low $\Gamma_j$, the main component is internal photons and the peak of neutrino fluence is lower than PeV, where the atmospheric background is stronger. 
Thus, the value of $\bar{N}_{\nu_\mu}$ is lower for a lower $\Gamma_j$ in the Southern sky although the peak fluence is higher as $\phi_{\nu_\mu}\propto \Gamma_j^{-2}$.
The probability of detecting more than one neutrino is given by
\begin{equation}
p_{n \geq 1} (\delta) = 1- \mathrm{exp}(-\bar{N}_{\nu_\mu}).
\end{equation}
This probability depends on $\delta$, and we calculate the probability averaged over the solid angle by \footnote{The probability differs from $p_{n\ge1}(\langle\bar{N}_{\nu_\mu}\rangle_{\rm ang})=1- \mathrm{exp}[- \int d\Omega \bar{N}_{\nu_\mu}/(4\pi)]$. }
\begin{equation}
    \langle p_{n \geq 1} \rangle _\mathrm{ang}
    = \frac{1}{4\pi} \int d\Omega \; p_{n \geq 1} (\delta).
\end{equation}
We calculate $\langle p_{n \geq 1} \rangle _\mathrm{ang}$ for various $t_\mathrm{dur}$ by using Eq. (\ref{eq:Lx}), and find that $\langle p_{n \geq 1} \rangle _\mathrm{ang}$ highly depends on $t_\mathrm{dur}$ as shown in \fg\ref{fig:p-t}.
There is a jump at $t_\mathrm{dur} = 10^2$ s for $\Gamma_j $ = 2000, caused by the condition of no cocoon photons, $R_\mathrm{coc} = \beta_{\rm{coc}} c t_\mathrm{dur} < r_\mathrm{diss}$.
The jump is an artifact because the number density of the cocoon photons in the dissipation region should change more smoothly. 
The bump at $t_\mathrm{dur} \sim 200$ s for $\Gamma_j=200 $ is caused by the shielding of cocoon photons due to the high optical depth. 
For $t_\mathrm{dur}>100$ s, the expansion of the cocoon decreases the cocoon temperature and the photon number density, leading to a low neutrino fluence for a high value of $t_{\rm dur}$.
For $t_\mathrm{dur}<100$ s, the internal photons dominate because $r_{\rm diss}>R_{\rm coc}$ is satisfied, and $\langle p_{n \geq 1} \rangle _\mathrm{ang}$ is high for a low value of $t_{\rm dur}$ due to $L_{X,\mathrm{iso}} \propto (t_\mathrm{dur}/10^2\ \mathrm{s})^{-2.5}$.

\subsection{Prospects for detecting neutrinos associated with GWs}
We estimate the expected number of neutrino detection associated with GWs, taking into account the distribution of $t_\mathrm{dur}$ and $L_{X,\mathrm{iso}}$.
We assume that the distribution of $t_\mathrm{dur}$ is lognormal:
\begin{equation}
F(t_\mathrm{dur}) = \frac{d N_{t_\mathrm{dur}}}{d\mathrm{log}_{10} \, (t_\mathrm{dur})} = F_0 \,\mathrm{exp}\left(-\frac{(\mathrm{log}_{10} \, (t_\mathrm{dur}/t_\mathrm{dur,0}))^2}{2 \sigma_{\mathrm{log}_{10} t_\mathrm{dur}}^2}\right),
\end{equation}
where $t_\mathrm{dur,0}$ and $\sigma_{\mathrm{log}_{10} t_\mathrm{dur}}^2$ are the mean and variance of the duration, respectively.
We fit the data of extended emission listed in \cite{Kisaka2017ApJ...846..142K} to obtain the values of $t_{\rm dur,0} = 10^{2.4}$ s and $\sigma_{\mathrm{log}_{10} t_{\rm dur}}^2 = 8.3\times10^{-2}$.
%, which are listed in \tb\ref{table2}.

The probability of detecting neutrinos from a sGRB is given by
\begin{equation}
P_{n \geq 1}  =  \int d(\mathrm{log}_{10} \, t_\mathrm{dur}) F(t_\mathrm{dur})  \, \langle p_{n \geq 1} \rangle _\mathrm{ang}.
\end{equation}

\begin{figure}
\gridline{\fig{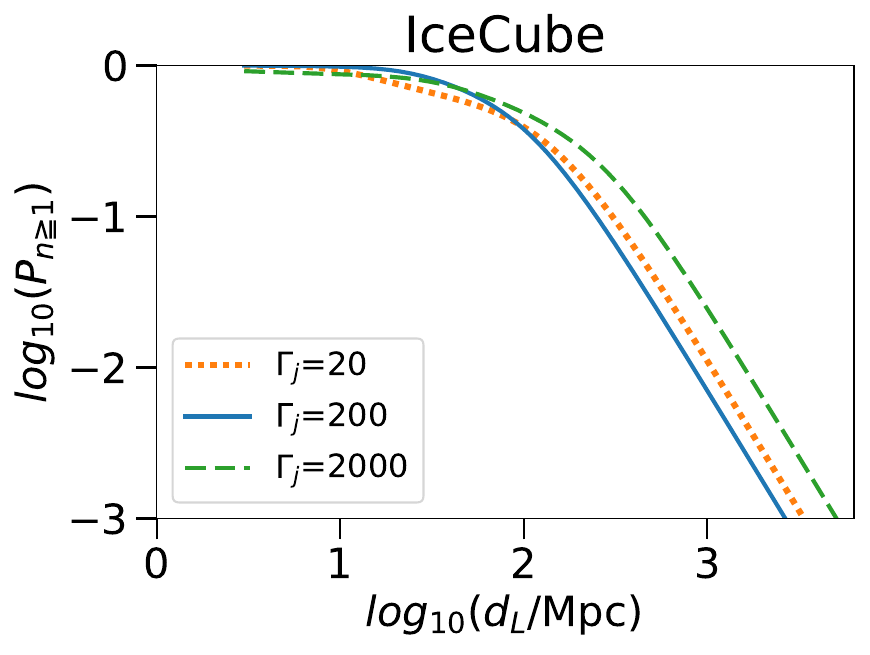}{0.4\textwidth}{}
}
\gridline{\fig{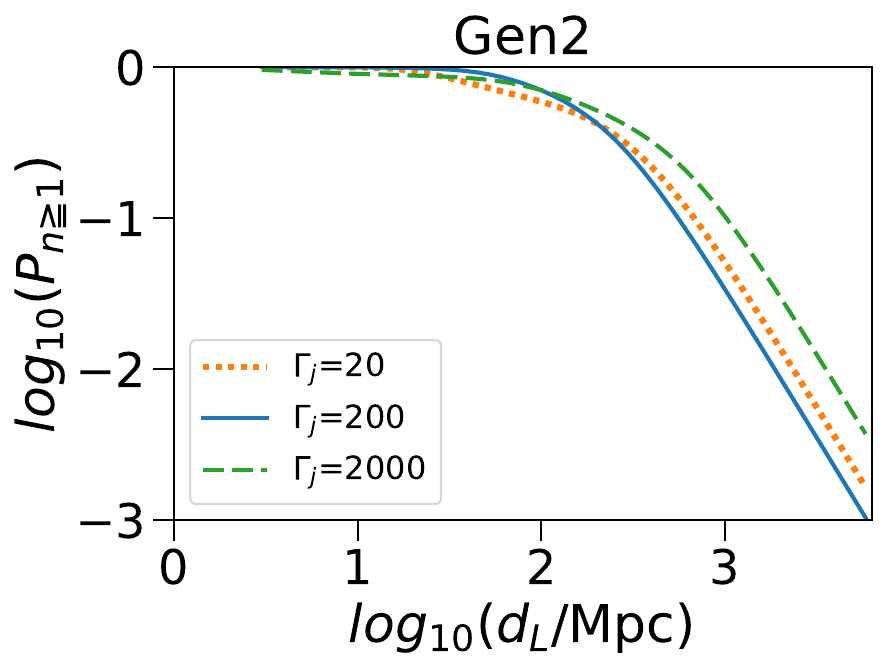}{0.4\textwidth}{}
          }
\caption{The probability of detecting more than one neutrinos as a function of $d_L$. The upper and lower panel are models for IceCube and IceCube-Gen2, respectively. The dotted, solid, and dashed lines represent probability for $\Gamma_j = 20$, 200, and 2000, respectively.     
\label{fig:p-dL}}
\end{figure}

We show dependence of $ P_{n \geq 1} $ on $d_L$ in \fg\ref{fig:p-dL}.
The behavior of the line for each $\Gamma_j$ is roughly the same for each $d_L$.
$P_{n \geq 1} \propto d_L^{-2}$ for high $d_L$ is consistent with $\phi_{\nu_\mu} \propto d_L^{-2}$ for a fixed luminosity.
Assuming the uniform distributions of the location and timing of sGRBs, we can estimate the probability of detecting more
than one neutrino associated with GW signal in $T_\mathrm{op}$ (yr) to be
\begin{equation}
q(T_\mathrm{op})  = 1- \mathrm{exp}\left(-T_\mathrm{op} R_\mathrm{sGRB} 4 \pi \int^{300 \mathrm{Mpc}} d(d_L) d_L^2 P_{n \geq 1}\right),
\end{equation}
where $R_\mathrm{sGRB}= 8\ \mathrm{Gpc}^{-3}\ \mathrm{yr}^{-1}$ is the event rate of sGRB \citep[e.g.,][]{Coward2012MNRAS.425.2668C}.

\begin{figure}
\gridline{\fig{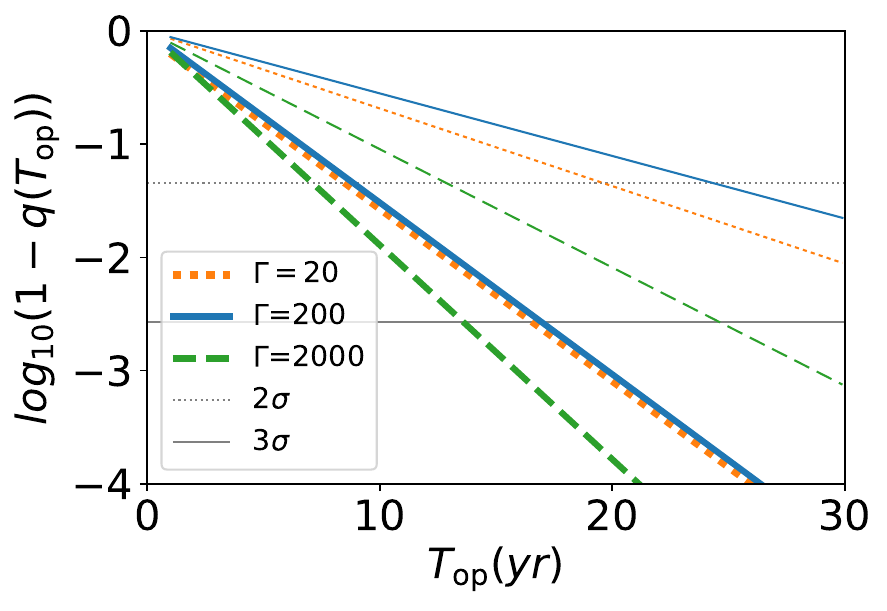}{0.4\textwidth}{(a) : dependence on $\Gamma_j$}
}
\gridline{\fig{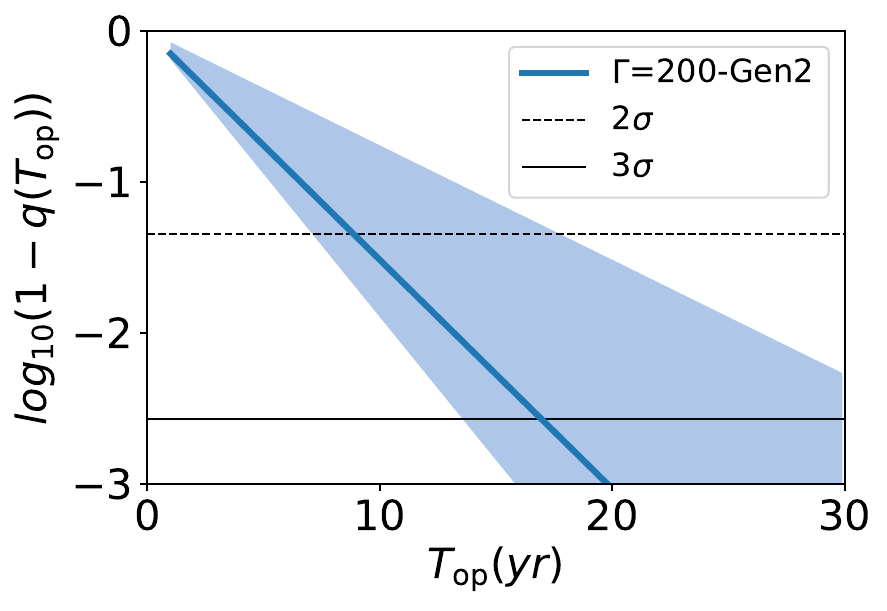}{0.4\textwidth}{(b) : dependence on $R_{sGRB}$}
          }
\caption{The probability of neutrino detection against the operation time.  The thin horizontal lines correspond to the significance of 2$\sigma$ and 3$\sigma$. (a) Thick and thin lines represent the models for IceCube and IceCube-Gen2, respectively. The dotted, solid, and dashed lines represent probability for $\Gamma_j = 20$, 200, and 2000, respectively. 
(b) The uncertainty for the local sGRB rate. The colored region corresponds to the probability between the cases with $R_{\rm sGRB}=4 \ \mathrm{Gpc}^{-3}\ \mathrm{yr}^{-1}$ and with $R_{\rm sGRB}=10\ \mathrm{Gpc}^{-3}\ \mathrm{yr}^{-1}$. \label{fig:q-T}}     
\end{figure}

The relation between $q$ and $T_\mathrm{op}$ for IceCube (the thick lines) and for Ice-Cube-Gen2 (the thin lines) are shown in \fg\ref{fig:q-T}. We can constrain our fiducial parameter set for 2$\sigma$ (3$\sigma$) confidence level within $\sim10$ ($\sim15$) years of operation by IceCube-Gen2, although it would take more than 25 years by IceCube even for the most optimistic case.
This indicates the importance of IceCube-Gen2 for detecting neutrinos from sGRBs and revealing the characteristics of prolonged jets.

We set the maximum luminosity distance as 300 Mpc, taking into account the sensitivity limit of BNS mergers for the second-generation GW detectors.
%O5 observing plan of GW. 
Then, $R_\mathrm{sGRB}$ should be the rate of local sGRB though it has large uncertainty due to the low number of the local sGRBs \citep{Wanderman2015MNRAS.448.3026W,Escorial2022arXiv221005695R}.
The bottom panel of \fg\ref{fig:q-T} shows that the result is sensitive to the rate, but the possibility becomes equivalent to
2$\sigma$ within 20 years of operation.
We need to determine the local sGRB rate to make a more solid prediction for neutrino detection.
In 20 years, the development of GW observatories will enable us to observe BNS mergers far from more than 300Mpc.
This will dramatically increase the GW events, which possibly leads to an earlier detection of a neutrino associated with a GW event. 

\section{GRB 211211A}
\label{sec:GRB211211A}
GRB 211211A is a long GRB whose prompt emission lasts for 13 s, followed by a soft extended component with a duration of 55 s \citep{Yang2022grb211211a}.
The progenitor of the GRB is, however, thought to be a merger of compact objects. Its host galaxy candidate is located at $d_L\simeq350$ Mpc, and the GRB occurred at the outskirts of the galaxy.
The prompt burst exhibits typical observational features of sGRBs, such as the negligible temporal lag, short variability timescale, and the position of the $\varepsilon_{\gamma,\rm pk}-E_{\rm iso}$ relation \citep{Troja2022Natur.612..228T}.
The tentative evidence of kilonova associated with the GRB also supports the BNS merger origin \citep{Rastinejad2022Natur.612..223R}.

Additionally, GeV gamma-rays are observed with the GRB in $10^3-10^4$ s after the prompt emission \citep{Mei2022Natur.612..236M}.
The gamma rays are explained by the external inverse Compton scattering process by non-thermal electrons accelerated in a prolonged jet (see \citealt{Zhang2022GeV_grb211211a} for another interpretation).
Although the GeV signature is observed at a much later phase than the extended emission, the same system might be realized during the extended emission phase.

Neutrino associations with the GRB have not been reported by gamma-ray follow-up (GFU) search of IceCube. 
This implies that we can obtain an upper limit of neutrino flux from the GRB using the effective area of GFU \citep{Aartsen2017APh....92...30A}. 

In Section \ref{Calculating the Spectrum}, we show that neutrino emission with plateau emission is not detectable by IceCube and IceCube-Gen2. 
This is consistent with the non-detection of neutrinos.
On the other hand, it is expected that neutrinos could be found with extended emission because the luminosity of the extended emission is much higher than the plateau emission. 
To check the consistency of our model, we calculate $\bar{N}_{\nu_\mu}$ from GRB 211211A based on our scenario.

\begin{figure}
 \centering
 \includegraphics[width = 0.49\textwidth]{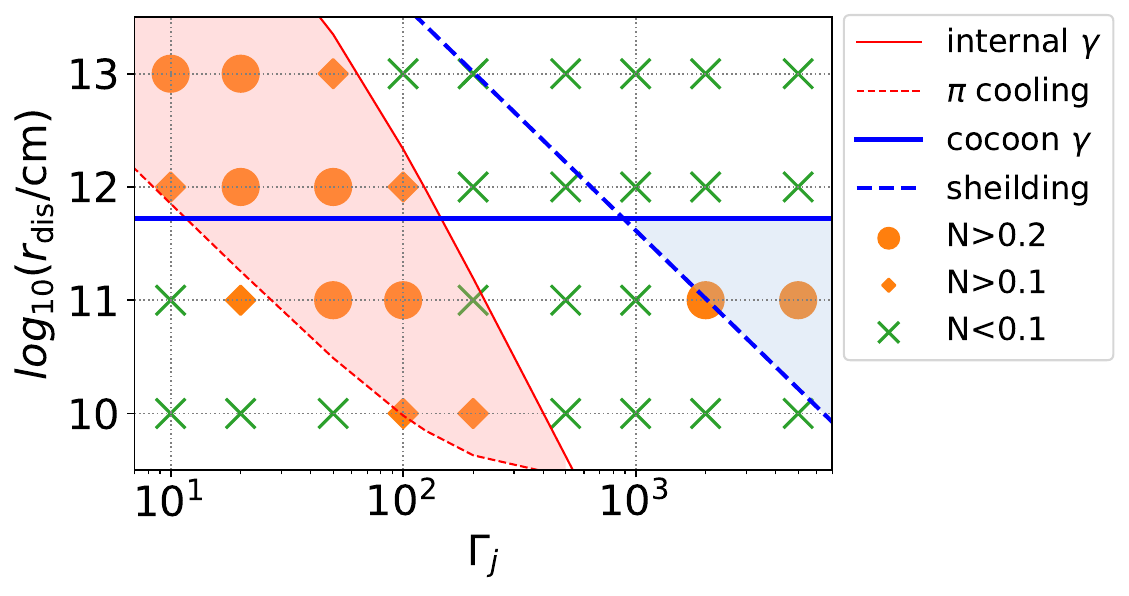}
 \caption{The chart of neutrino number from GRB 211211A. 
 The orange circles, orange diamonds, the green crosses indicate the parameters for $0.2<\bar{N}_{\nu_\mu}$, $0.1<\bar{N}_{\nu_\mu}<0.2$, and $\bar{N}_{\nu_\mu}<0.1$, respectively.
 The red-solid and blue-solid lines indicate the boundary for efficient pion production by internal and cocoon photons, respectively. The red-dashed line is the boundary for efficient pion cooling, and the blue-dashed line is the boundary for the shielding of the cocoon photons.
 Red-shaded and blue-shaded regions are parameter space where neutrinos are efficiently produced by internal and cocoon photons, respectively.
}
 \label{fig:GRB211211A}
\end{figure}

The property of the event and its extended emission is summarized in \tb\ref{GRB211211A} (see Table 1 of \citealt{Yang2022grb211211a} for other quantities).
Since the duration of the extended emission is relatively short, the dissipation radius should be small, $r_\mathrm{diss} \lesssim 5.3 \times10^{11}$ cm, for the cocoon photons to enter into dissipation regions.
The calculations are performed by setting $\Gamma = 10-5000$ and $r_\mathrm{diss} = 10^{10}-10^{13}$ cm though they include extreme parameters.
The results are shown in \tb\ref{GRB211211A_result} and \fg\ref{fig:GRB211211A}.
These indicate that the expected numbers of neutrinos observed by IceCube are less than 1 for all the parameter sets.
The expected number for GFU is also less than 1 because the effective area of GFU  is smaller than that of the point source in the northern hemisphere ($\delta >0$).
These results are consistent with the absence of GFU alerts associated with GRB 211211A.
We cannot constrain the parameter space with the current facilities.

The observation by IceCube-Gen2 is important.
The circles in \fg\ref{fig:GRB211211A} show parameters where the calculated expected numbers of neutrinos by IceCube are greater than 0.2.
The effective area of IceCube-Gen2 is five times larger than that of IceCube, and IceCube-Gen2 will be able to detect a neutrino from GRB 211211A-like bursts for those parameters or put constraint on 68\% confidence level.

The neutrino signals from GRB 211211A are stronger for low $\Gamma_j$ region owing to the intense internal photon field. The neutrino signals can be stronger for a medium $r_\mathrm{diss}$ and high $\Gamma_j$ region because of the cocoon photons. 
These results can be analytically understood by the following discussion.

First, we discuss the conditions when the internal photons are dominant. For efficient neutrino production, the pion production efficiency needs to be high.
Considering the large effective area of IceCube around $\varepsilon_{\nu_\mu}\sim100$ TeV,
the condition of efficient pion production is determine by $(t_\mathrm{ad}^{\prime}/t^{\prime}_{p\gamma,\mathrm{int}})|_{\varepsilon_\nu = 100\ \mathrm{TeV}} \geq 1$.
This condition can be represented by
\begin{equation}
\begin{split}
 r_\mathrm{dis,12} \leq 6.0 &\  \Gamma_{j,2}^{-2}L_{\gamma,\mathrm{iso},50}  \left(\frac{\varepsilon_{\gamma,\mathrm{pk}}}{100\ \mathrm{keV}}\right)^{-1}\\
    &\times \left[1+0.54\ \left(\frac{\varepsilon_{\gamma, \mathrm{pk}}}{100\ \mathrm{keV}}\right)\ \Gamma_{j,2}^{2}\right]^{-1},
\end{split}
 \label{eq;fast_pi_pro_int}
\end{equation}
which is derived from Equation (28) in \cite{Kimura2022arXiv220206480K}. 
This condition is shown by red-solid line in \fg\ref{fig:GRB211211A}.
In addition, the pion cooling needs to be inefficient. 
The condition of the inefficient pion cooling is determined by $\varepsilon_{\nu,\mathrm{cool}}\geq 100$ TeV,
where $\varepsilon_{\nu,\mathrm{cool}}$ is defined by $(t^{\prime}_{\pi,\mathrm{cool}}|_{\varepsilon^\prime_p = 20 \varepsilon_{\nu,\mathrm{cool}}/\Gamma_j}/t^{\prime}_\mathrm{\pi,dec}|_{\varepsilon^\prime_p = 20 \varepsilon_{\nu,\mathrm{cool}}/\Gamma_j}) =1$.
The condition leads to 
\begin{equation}
 r_\mathrm{dis,12} \geq 4.6\times10^{-3}\  \Gamma_{j,2}^{-2}L_{\gamma,\mathrm{iso},50}^{1/2}\left(\frac{\xi_\mathrm{B}}{0.33}\right)^{1/2} +  2.5\times10^{-3}.
 \label{eq;fast_pi_cool}
\end{equation}
The first term of the right-hand-side is the contribution by the synchrotron cooling and the other is by the adiabatic cooling.
The red-dotted line represents this condition.

Next, we consider the case where the cocoon photons are dominant.
For efficient neutrino production, the cocoon needs to cover the dissipation region; $r_\mathrm{diss}\leq R_\mathrm{coc}$, which is shown as the blue-solid line.
If this condition is satisfied, the cocoon photons always lead to the efficient neutrino production, i.e., $(t^\prime_\mathrm{ad}/t^{\prime}_{p\gamma,\mathrm{coc}})|_\mathrm{peak} \geq 1$.
The cocoon photons are shielded if the lateral optical depth, $\tau$, is large.
The strong shielding can be avoided if $e^{-\tau} \geq 1/10$ is satisified.
This condition can be rewritten as
\begin{equation}
 r_\mathrm{dis,12} \geq  40\times \Gamma_{j,2}^{-2} L_{\gamma,\mathrm{iso},50} \times \ln10.
 \label{eq;shielding}
\end{equation}
This condition is shown in the blue-dashed line.

In summary, the neutrino production is efficient below the solid lines and above the dashed lines in \fg\ref{fig:GRB211211A}. In the red-colored region, the neutrinos are produced by the internal photons, whereas the cocoon photons are important in the blue-colored region.  

\begin{table}
\caption{Properties of GRB 211211A}
\centering
\begin{tabular}{cccc}
\hline \hline
RA&Dec (= $\delta$)\\
\hline
14h 09m 05s&+27$^\circ$ 53' 01"\\
\hline \hline
 $t_\mathrm{dur}$& $\varepsilon_{\gamma,\mathrm{pk}}$& $\alpha$ & $\beta$ \\
 (s)&  (keV)  &  &  \\
 \hline
 55 & 82 & $-0.97$&$-2.02$   \\
\hline \hline 
   Energy fluence   & $L_{X,\mathrm{iso}}$& $\varepsilon_{\mathrm{X,min}}$, $\varepsilon_{\mathrm{X,max}}$& $d_L$\\
  (erg/$\mathrm{cm}^2$)    & (erg/s)&  (keV)& (Mpc)\\
  \hline
  $1.6\times 10^{-4}$   & $4.3\times 10^{49}$& 15, 150 (BAT)& 350\\

 \label{GRB211211A}
 \end{tabular}
\end{table}

\begin{table}
\caption{Expected neutrino number from GRB 211211A}
\centering
\begin{tabular}{rl|ccc}
\hline \hline
 & $\Gamma_j$ & 20 & 200 & 2000\\
 $r_{\mathrm{dis}}$(cm) &  & & \\
\hline 
$10^{13}$ &  & 4.1$\times10^{-1}$&5.3$\times10^{-3}$ &9.1$\times10^{-7}$ \\
$10^{12}$ &  & 3.0$\times10^{-1}$&2.7$\times10^{-2}$ &6.0$\times10^{-6}$ \\
$10^{11}$ &  & 1.2$\times10^{-1}$&7.5$\times10^{-2}$ &8.2$\times10^{-1}$ \\
$10^{10}$ &  & 2.1$\times10^{-2}$&1.0$\times10^{-1}$ &1.1$\times10^{-4}$ \\
\label{GRB211211A_result}
\end{tabular}
\end{table}

\section{Summary and Discussion}
\label{sec:Summary}
We calculated the neutrino fluence emitted by the prolonged jet of sGRB, which can be associated with GW events. We take into account the photons entering into the jet from the cocoon formed by the jet-ejecta interaction.
Owing to the contribution by the cocoon photons, the peak neutrino fluence has a weak dependence on the Lorentz factor of the jet, $\Gamma_j$.
We found that the peak fluence is approximately $5.4\times10^{-3}$ GeV $\mathrm{cm}^{-2}$ at $\varepsilon_{\nu} \sim$PeV from a single sGRB with our fiducial parameters shown in \tb\ref{fiducial parameters}. 
We expect that the expected number of the neutrino event will be 0.1 by IceCube, while it will be 0.5 by IceCube-Gen2. 
Assuming the homegeneous spatial distribution and the log-normal duration distribution of sGRBs, we found that the possibility of neutrino detection associated with GW events is sufficiently high if we continue to observe cosmic neutrinos with IceCube-Gen2 for 10 years. 
Even if the future observation results in no detection of neutrinos associated with GWs, we can put a strong constraint on the parameter space of the late-time jets.
We also applied our model to GRB 211211A. We found that no associated neutrino signal in IceCube with GRB 211211A is consistent with our model, but we can put a meaningful constraint with future GRB 211211A-like events if IceCube-Gen2 is in operation.

We assumed that CRs are accelerated at the dissipation radius, but we need to be cautious about the condition for CR production.
In the internal shock model, the radiation mediated shock can be formed if the upstream of the shock is optically thick, 
which prevents CRs from being accelerated \citep{Murase2013PhRvL.111l1102M,Kimura2018PhRvD..98d3020K}. 
This effect can drastically reduce neutrino fluence. 
However, our scenario can avoid the condition even for $\Gamma_j=20$ as long as the Lorentz factor of the shock upstream is as high as 100.

In addition, we should note that the energy density of the internal photon, $\int^{\varepsilon_{\mathrm{X,Max}/\Gamma_j}}_{\varepsilon_{\mathrm{X, min}/\Gamma_j}} d\varepsilon^{\prime}_{\gamma} \varepsilon^{\prime}_{\gamma}({dn^{\prime}_\gamma}/{d\varepsilon^{\prime}_{\gamma}} ) = L_{\mathrm{X, iso}}/(4 \pi\Gamma_j^2 r_\mathrm{diss}^2 c) $, can be modified if $\tau_j>1$ is satisfied.
The internal photon density can be constant in radius if a part of the kinetic energy of the jet dissipates inside the photosphere, as discussed in \cite{Rees2005ApJ...628..847R}. In this case, our results are not affected.
However, the photon luminosity at the dissipation region can be higher than that at the photosphere if the dissipation cannot compensate the energy loss by the adiabatic expansion. In this case, the internal photon density at the dissipation region is higher than that at the photosphere.
Nevertheless, this effect has little influence on the neutrino fluence because of the efficient neutrino production, $f_{p\gamma} =1 $, for the cases with $\tau_j\gtrsim1$. 
Therefore, our conclusions should be unchanged as long as CRs are accelerated at the given $r_{\rm diss}$.

We can constrain other parameters, such as dissipation radius $r_{\rm diss}$, more strongly than the previous study if we ignore the dependence on $\Gamma_j$.
The case with $r_\mathrm{diss}>10^{12}$ cm will favor the internal shock model or  Internal Collision-induced Magnetic Reconnection and Turbulence (ICMART; \citealt{Zhang2011ApJ...726...90Z,Zhang2013PhRvL.110l1101Z}), while the case with $r_\mathrm{diss}<10^{12}$ cm would favor the dissipative photosphere scenario \citep{Rees2005ApJ...628..847R}.

The detection of neutrinos from the sGRBs will be a smoking-gun signature of the hadronic CR acceleration in the sGRB environment. 
This can constrain the composition of the jet in extended emission, i.e., whether the prolonged jets are leptonic or baryonic.
Baryonic jets demand a baryon injection process into jets, such as neutron diffusion from neutron-rich matter \citep{Belovobodov2003neutron,Levinson2003neutron}, 
which is thought to be the baryon injection process into the prompt jets.
In the phase of the late-time emission, the mass accretion rate onto the remnant object might be too low to form a neutron-rich material \citep[e.g.,][]{Kohri2005ndaf}.
Magnetar models are often discussed as a possible explanation of extended emissions only by the leptonic process \citep{Dai1998magnetar}.
Neutrino observations will potentially be able to rule out some leptonic models, which will clarify the jet launching mechanism.

\cite{abbasi2022grbIceCube} concludes that time-integrated fluences in 300 s after the event for all northern sGRBs (183 events) are less than $1\times10^{-1}$ GeV $\mathrm{cm}^{-2}$ in total by analyzing the correlation of the GRB catalog detected by \textit{Swift} or \textit{Fermi} and IceCube data sets.
This means that the upper limit of fluence from a single sGRB is about $5 \times10^{-4}$ GeV $\mathrm{cm}^{-2}$ on average.
The data include mostly the signals of sGRBs from cosmological distance, which is not the focus of our study.
If we assume that the redshifts of the sGRBs analyzed in \cite{abbasi2022grbIceCube} are typically about $z\sim0.5$  (corresponding to $d_L \sim 3$ Gpc), the upper limit of neutrino luminosity becomes about $3\times10^{48}\rm~erg~s^{-1}$.
The neutrino luminosity in our model, $\sim 5\times10^{47}\rm~erg~s^{-1}$, is lower than the upper limit and consistent with the IceCube data.

The high-energy gamma-ray observations also provide a test for our model.
The astrophysical neutrinos are inevitably accompanied by gamma-rays produced by $\pi^0$ decay with the similar energy and luminosity, 
because the reaction rate of $\pi^0$ production is almost the same as that of $\pi^\pm$ production.
However, the peak energy of gamma-rays is not 100 TeV - PeV, because they are to cascade into lower energies by the interaction with low-energy photons. The resulting peak of the escaped photon spectrum is expected to be MeV -- GeV.
Gamma-rays in this energy band will be detected by
future telescopes, such as eASTROGAM \citep{DeAngelis2017ExA....44...25D}, GRAMS \citep{Aramaki2020APh...114..107A}, and AMEGO-X \citep{Caputo2022arXiv220804990C}.
The multi-messenger approach, using gamma-ray and the neutrino signals, will be a powerful tool to constrain the physical conditions of the dissipation region.

Neutrino detection associated with GWs enables us to probe the phenomena that cannot be investigated only by EM signals.
Choked jet systems, where jets fail to break out from the stellar envelope or kilonova ejecta, are a good example \citep{Murase2013PhRvL.111l1102M,Kimura2018PhRvD..98d3020K}.  
The threshold timescale in the duration distribution of prompt emissions of sGRBs indicates the existence of such events \citep{Moharana2017MNRAS.472L..55M}.
\cite{Matsumoto2018ApJ...866L..16M} showed that prolonged jet can break out from the ejecta even if prompt jet fails to penetrate it.
In such a scenario, we cannot observe the emission from prompt jets but can observe the extended emission from the prolonged jets.
X-ray observations found a few candidates for such a delayed breakout event \citep{Xue2019Natur.568..198X}, but it is difficult to confirm.
Based on our model, we expect detection of neutrinos and GWs from the delayed breakout event. Thus, GW-neutrino association without prompt gamma rays would strongly support the scenario of the delayed breakout. 
The event rate of the choked jets is expected to be approximately 0.4 times lower than that of the successful sGRBs \citep{Sarin2022PhRvD.105h3004S}, although it can be increased by the uncertainty of the event rate of sGRBs and BNS mergers. 
If the fraction of the choked jet is on the high end, the delayed breakout will be a major component, and we will be able to detect much more GW-neutrino association.
Thus, the GW-neutrino association can be a powerful tool to probe the central engine activity.

\begin{acknowledgments}
This work is supported by Graduate Program on Physics for the Universe (GP-PU), Tohoku University (R.M.), JSPS KAKENHI No. 22K14028 (S.S.K.) and 18H01245 (K.T.). This work of K.M. is supported by the NSF Grant No.~AST-1908689, No.~AST-2108466 and No.~AST-2108467, and KAKENHI No.~20H01901. 
S.S.K. acknowledges the support by the Tohoku Initiative for Fostering Global Researchers for Interdisciplinary Sciences (TI-FRIS) of MEXT's Strategic Professional Development Program for Young Researchers.
\end{acknowledgments}

\appendix
\section{Analytic Explanation of Neutrino Spectra}
\label{App_a}
In this section, we analytically estimate the neutrino fluence when the cocoon photons dominate over the internal photons.
If we assume $g(\varepsilon_{\pi},\varepsilon_{\nu_\mu}) = \delta(\varepsilon_{\pi}/4-\varepsilon_{\nu_\mu})$ and $g(\varepsilon_{\mu},\varepsilon_{\nu_e}) = \delta(\varepsilon_{\mu}/3-\varepsilon_{\nu_e})$, the neutrino spectra are given by

\begin{equation}
\varepsilon_{\nu_\mu}^2\frac{dN_{\nu_\mu}}{d\varepsilon_{\nu_\mu}} \approx \frac{1}{8} f_{p\gamma}f_{{\rm sup},\pi} \varepsilon_p^{2}\left.\frac{dN_p}{d\varepsilon_{p}}\right|_{\varepsilon_{p} = 20\varepsilon_{\nu_\mu}}
\label{eq;A1}
\end{equation}
\begin{equation}
\varepsilon_{\bar{\nu}_\mu}^2\frac{dN_{\bar{\nu}_\mu}}{d\varepsilon_{\bar{\nu}_\mu}} \approx \varepsilon_{\nu_e}^2\frac{dN_{\nu_e}}{d\varepsilon_{\nu_e}} \approx  \frac{1}{8} f_{p\gamma}f_{{\rm sup},\pi}f_{sup,\mu} \varepsilon_p^{2}\left.\frac{dN_p}{d\varepsilon_{p}}\right|_{\varepsilon_{p} = 20\varepsilon_{\nu_\mu}}.
\label{eq;A2}
\end{equation}
These are useful to understand the resulting spectrum analytically \citep[e.g.,][for a review]{Kimura2022arXiv220206480K}.
This expression assumes that all the decay products produced by a pion with $\varepsilon_{\pi}$ share the same amount of energy: $\varepsilon_{\nu_\mu} \approx \varepsilon_{\nu_e}\approx \varepsilon_{\bar{\nu}_\mu}\approx \varepsilon_e\approx \varepsilon_\pi/4$.

To estimate $f_{p\gamma}$ and the fluence, we can roughly estimate $t^{\prime}_{p\gamma}$ by using additional approximations.
Considering only cocoon photons, which are the dominant component for the fiducial parameters, and $\sigma\kappa \sim \sigma_{0,\mathrm{eff}} \theta(\bar{\varepsilon}_{\gamma} - \bar{\varepsilon}_\mathrm{th} )$ as an approximation, we can perform the calculation of \equ\eqref{pg}. 
Using Eq.(69) of \cite{Dermer2012ApJ...755..147D}, we can obtain timescale of photomeson production with the cocoon photons as
\begin{eqnarray}
t^{\prime-1}_{p\gamma,\mathrm{coc}} = & \frac{16\pi k_B^3 T^3_\mathrm{coc} \sigma_{0,\mathrm{eff}}}{h^3c^2}\Gamma_j  \int^{\infty}_{\tilde{\varepsilon}^{\prime}_p/\varepsilon^\prime_p}
dy \  y\ \mathrm{ln}(1-e^{-y})^{-1} \nonumber\\ 
\sim & \frac{16\pi k_B^3 T^3_\mathrm{coc} \sigma_{0,\mathrm{eff}}}{h^3c^2}\Gamma_j \left\{
\begin{array}{ll}
  \zeta(3) & (\varepsilon^{\prime}_p \gg \tilde{\varepsilon}^{\prime}_p)\\
  \tilde{\varepsilon}^{\prime}_p/\varepsilon^\prime_p \  e^{-\tilde{\varepsilon}^{\prime}_p/\varepsilon^\prime_p} & (\varepsilon^{\prime}_p \ll \tilde{\varepsilon}^{\prime}_p),
\end{array}
\right.
\end{eqnarray}
where $\tilde{\varepsilon}^{\prime}_p = (\bar{\varepsilon}_\mathrm{th} m_pc^2)/(2 k_B T_{\mathrm{coc}} \Gamma_j)$ is the threshold energy of protons for photomeson production with typical cocoon photons,
and $\zeta(x)$ is the zeta function. 
The lightblue-thin-dotted-dashed lines shown in \fg\ref{fig:tp-e}, which is almost completely overlapped with the thick-dot-dashed line in our fiducial model, can be well explained by this approximate formula .
$t^{\prime-1}_{p\gamma,\mathrm{coc}}$ is constant for $\varepsilon^\prime_p\gtrsim \tilde{\varepsilon}^{\prime}_p$, and it has an exponential cutoff at lower energies.
The cutoff neutrino energy in the observer frame is estimated to be
\begin{equation}
\varepsilon_\nu \sim  \frac{\tilde{\varepsilon}^{\prime}_p}{20\Gamma_j} = \frac{\bar{\varepsilon}_\mathrm{th} m_pc^2}{40 k_B T_{\mathrm{coc}}}\sim 0.13 \left(\frac{\bar{\varepsilon}_\mathrm{th}}{0.1\ \rm{GeV}}\right)\left(\frac{k_B T_\mathrm{coc}}{20\ \rm{eV}}\right)^{-1}\  \rm{PeV}    
\label{eq:enucut}
\end{equation}
which does not depend on $\Gamma_j$.

For $\varepsilon'_p>\tilde{\varepsilon}^{\prime}_p$, we can write
\begin{equation}
t^{\prime-1}_{p\gamma,\mathrm{coc}} \sim 88 \ \left(\frac{\sigma_{0,\mathrm{eff}}}{6\times 10^{-29}\ \rm{cm}^2}\right)\left(\frac{k_B T_\mathrm{coc}}{20\ \rm{eV}}\right)^3\left(\frac{\Gamma_j}{200}\right)\  \ \rm{s}^{-1}.
\label{eq;a4}
\end{equation}
Based on \fg\ref{fig:tp-e}, the photomeson production with the cocoon photons is the most efficient process, and all the protons of $\varepsilon'_p>\tilde{\varepsilon}^{\prime}_p$ lose their energy.
Thus, the peak fluence contributed by cocoon photons does not depend on $\Gamma_j$.
Even for the case with $t^{\prime}_{\rm ad}< t^{\prime}_{p\gamma,\rm coc}$, both of the timescales have the same $\Gamma_j$ dependence, and therefore, the pion production efficiency and the neutrino fluence do not depend on $\Gamma_j$ as long as the cocoon photons are dominant for neutrino production.

The high-energy cutoff in the neutrino spectra is caused by the proton injection spectrum.
The cutoff energy of protons in the observer frame depends on $\Gamma_j$.
In the range, $t^{\prime-1}_\mathrm{cool} \sim t^{\prime-1}_{p\gamma,\mathrm{coc}} \propto \Gamma_j$ is satisfied and $t^{\prime-1}_{\mathrm{acc}}$ is proportional to $\Gamma_j^{-1}$ due to Lorentz transformation of magnetic field.
This leads to $\varepsilon^\prime_{p,\mathrm{cut}}\propto \Gamma_j^{-2}$ and $\varepsilon_{p,\mathrm{cut}} \propto \Gamma_j^{-1}$.
The cutoff energy is $\varepsilon^\prime_{p,\mathrm{cut}} \sim 8.2 \times 10^6\ (\Gamma_j/200)^{-2}$ GeV for the proton energy in the comoving frame of the jet and $\varepsilon_{\nu} \sim 8.2 \times 10^7 \ (\Gamma_j/200)^{-1}$ GeV for neutrino energy in the observer frame for the fiducial parameters.

For $\varepsilon_{\nu_\mu}<100$ TeV, the resultant fluence shown in \fg\ref{fig:phi-e} is $\varepsilon_\nu^{2}({dN_\nu}/{d\varepsilon_{\nu}})\propto \varepsilon_\nu^{2}$, while $f_{p\gamma,\mathrm{coc}}\times \varepsilon_p^{2}({dN_p}/{d\varepsilon_{p}})$ should have a cutoff feature.
This is because of the different treatment of $g(\varepsilon_{i},\varepsilon_{j})$.
The cutoff feature of the lower energy is not realistic since the decaying pions with the peak energy must produce neutrinos with energies lower than $(1/4)\varepsilon_{\pi}$.

%% For this sample we use BibTeX plus aasjournals.bst to generate the
%% the bibliography. The sample631.bib file was populated from ADS. To
%% get the citations to show in the compiled file do the following:
%%
%% pdflatex sample631.tex
%% bibtext sample631
%% pdflatex sample631.tex
%% pdflatex sample631.tex

\bibliography{sample631}{}
\bibliographystyle{aasjournal}

%% This command is needed to show the entire author+affiliation list when
%% the collaboration and author truncation commands are used.  It has to
%% go at the end of the manuscript.
%\allauthors

%% Include this line if you are using the \added, \replaced, \deleted
%% commands to see a summary list of all changes at the end of the article.
%\listofchanges

\end{document}